\documentstyle[12pt]{article}
\topmargin - 1.5 cm % for epson - 3,5 cm. for laserjet appr. -10pt
\begin{document}

\begin{center}
{\large{\bf Dyon-Oscillator Duality. Hidden Symmetry
of the Yang-Coulomb Monopole}}
\end{center}
\begin{center}
{\bf Levon Mardoyan${}^1$}
\end{center}
\begin{center}
Yerevan State University\\
International Center for Advanced Studies\\
Alex Manougian str., 1, 375025 Yerevan, Armenia
\end{center}

\footnotetext[1]{E-mail: mardoyan@icas.ysu.am}

\begin{center}
\underline{Abstract}
\end{center}
{\small In this article, in the framework of an analytical approach
and with help of the generalized version of the Hurwitz transformation
the five--dimensional bound system composed of the Yang monople
coupled to a particle of the isospin by $SU(2)$ and Coulomb
interaction is constructed from the eight-dimensional quantum
oscillator. The generalized Runge-Lentz vector and the $SO(6)$
group of the hidden symmetry are established. It is also shown
that group of hidden symmetry makes it possible to calculate
the spectrum of system by a pure algebraic method.}

\section{Introduction}

The objective of the present work is to illustrate the property of the
Shr\"odinger equation which is called here the dyon--oscillator duality.
The property is in the following. The Sch\"odinger equation for an oscillator
possesses two parameters  -- the energy $E$ and the cyclic frequency
$\omega$. The quantization leads to the constraint $E=\hbar\omega(N+D/2)$,
where $N=0,1,2,\dots$, and $D$ is the dimension of  the configuration
space of the oscillator. If $\omega$ is fixed then $E$ is quantized,
and that is the standard situation. Imagine for a moment that now
$E$ is fixed. Whence, necessarily $\omega$ is quantized, and we are
in nonstandard situation. The question is whether the nonstandard situation
corresponds to any physics, i.e. is it possible  to find such a
transformation that converts the oscillator to some physical system with
coupling constant $\alpha$, being a function of $E$, and energy
$\varepsilon$, depending on $\omega$? If there  exists such a transformation,
we can confirm that the "nonstandard oscillator" is identical to that
physical system. Below will be shown the validity of the  described picture
for  dimensions $D=1,2,4,8$, and that the final system is bound system
of charge--particle (remind, that dyon is the hypothetical particle
introduced by Schwinger [1], which is unlike the Dirac monopole endowed with
not just  magnetic but  electric charge as well). As the "standard" and
"nonstandard" regimes are mutually exclusive, the  initial oscillator
and the final "charge--dyon" system are dual to each other, and that
explains the relevancy of the term "dyon--oscillator duality".
Note also, that  in the initial system the spectrum is discrete only,
i.e. the particle has just a finite motion (for such cases it is used to say
that we have a model with confinement). Generally speaking, the spectrum of
the final system includes the discrete spectrum as well as the continuous
one, i.e. in that model there is no confinement. However, unlike the first
model, in the second model we have monopoles. There is some analogy between
the dyon--oscillator and the Seiberg--Witten duality [2], according to which
the gauge theories with strong interactions are equivalent to the theories
having weak interaction on one hand and topological nontrivial objects,
such as monopoles and dyons are, on the other hand.

\section{Coulomb-Oscillator Duality}

Let us consider the equation
%===============================(1)=================================
\begin{eqnarray}
\frac{d^2 R}{d u^2}+\frac{D-1}{u}\frac{d R}{d u}-\frac{L(L+D-2)}{u^2}R
+\frac{2m}{\hbar^2}
\left(
E-\frac{m\omega^2 u^2}{2}
\right)R=0.
\end{eqnarray}
%=======================================================================
Here $R$ is the radial part of the wave function for $D$-dimensional
oscillator ($D > 2$), $L=0,1,2,\dots$ are the eigenvalues of the global
angular momentum.

After substitution $r=u^2$ the equation (1) transforms to the equation
%==================================(2)===============================
\begin{eqnarray}
\frac{d^2 R}{d r^2}+\frac{d-1}{r}\frac{d R}{d r}-\frac{l(l+d-2)}{r^2}R
+\frac{2m}{\hbar^2}
\left(\epsilon+\frac{e^2}{r}\right)R=0,
\end{eqnarray}
%====================================================================
where $d=D/2+1$, $l=L/2$,
%===================================(3)==============================
\begin{eqnarray}
\epsilon=-m\omega^2/8, \qquad  e^2=E/4.
\end{eqnarray}
%====================================================================
This is quite an unexpected result. If $D=4,6,8,\dots$, then
$d=3,4,5,\dots$, and the equation (2) is formally identical to the radial
equation for $d$-dimensional Coulomb problem (for odd $D > 2$ the value
of $d$ is half-integer and so cannot have the meaning of the dimension of the
space in the usual sense). Then, $l$ takes not only integer but half-integer
values as well, hence it has a meaning of general momentum and a question arise
about the origin of the  fermion degree of freedom. The answer to the
question will be given later. Finally, as was mentioned in  first section,
the equations (1) and (2) are dual to each other and the duality
transformation is $r=u^2$.

Up to now, just the radial part of the wave function of the oscillator has
been considered. For the Schr\"odinger equation we must take into account
the angular part as well. Thus, the duality transformation must also include
the transformation of angular variables. If we interpret the change of
variables $r=u^2$ as the mechanism of generation of electric charge,
(as will be shown later) the transformation of some angular variables is
responsible for the generation of magnetic charges.

Therefore, we return to the condition $r=u^2$. In the Cartesian coordinates
this condition has the form
%=================================================================
\begin{eqnarray*}
x_0^2 + x_1^2 + \cdots + x_{d-1}^2 = \left(u_0^2 + u_1^2 +
\cdots + u_{D-1}^2\right)^2,
\end{eqnarray*}
%====================================================================
which is called the Euler's identity. According to Hurwitz theorem [3]
this equation has a solution bilinear in $u_\mu$ only for the
following pairs of numbers
%=================================================================
\begin{eqnarray*}
(D, d) = (1, 1); (2, 2); (4, 3); (8, 5).
\end{eqnarray*}
%====================================================================

Three remarkable circumstances are connected with the last relation:

\begin{itemize}
\item (D, d) = (2, 2) -- is the Levi-Civita transformation [4]. \\
(D, d) = (4, 3) -- is the Kustaanheimo-Stiefel transformation [5]. \\
(D, d) = (8, 5) -- is the Hurwitz transformation [6,7,8].
\item The transformation (4) establishes the connection between two
fundamental problems of mechanics, the oscillator and Kepler problems.
\item The "magic" numbers $D=1,2,4,8$ have the direct relation to the
existence fact of four basic algebraic structures: real numbers, complex
numbers, quaternions and octanions.
\end{itemize}

Later on we consider only the case $(D, d) = (8, 5)$.

\section{Hurwitz Transformation}

We can write the solution of the Euler's indentity as
%===================================(4)==============================
\begin{eqnarray}
x = H(u; D) u,
\end{eqnarray}
%====================================================================
which can be interpreted as a bilinear transformation that maps one
Euclidean space into another. Here $D$ is the dimension of the space,
$H$ is the matrix $D\times D$ with the elements $u_\mu$, and $x, u$
are $D$-dimensional columns composed from $x_j, u_\mu$ and, possibly, zeroes.
So, for the Levi--Civita and Kustaanheimo--Stiefel transformations we have
%====================================================================
\begin{eqnarray*}
\left(
\begin{array}{c}
x_1\\x_2
\end{array}
\right)=\left(
\begin{array}{cc}
u_1&-u_2\\u_2&u_1
\end{array}
\right)
\left(
\begin{array}{c}
u_1\\u_2
\end{array}
\right),
\end{eqnarray*}
%====================================================================
%====================================================================
\begin{eqnarray*}
\left(
\begin{array}{c}
x_1\\x_2\\x_3\\0
\end{array}
\right)=\left(
\begin{array}{cccc}
u_3&-u_4&u_1&-u_2\\
u_4&u_3&u_2&u_1\\
u_1&u_2&-u_3&-u_4\\
u_2&-u_1&-u_4&u_3
\end{array}
\right)
\left(
\begin{array}{c}
u_1\\u_2\\u_3\\u_4
\end{array}
\right).
\end{eqnarray*}
%====================================================================
The matrixes $H(u;2)$ and $H(u;4)$ have the property
%====================================================================
\begin{eqnarray*}
H(u;2)\, H^T(u;2)=u^2 E(2),\quad
H(u;4)\, H^T(u;2)=u^2 E(4),
\end{eqnarray*}
%====================================================================
where $"T"$ means the sign of the transposition, $E(2)$ and $E(4)$
are the unit matrixes. Due to these properties the Euler identities are
fulfilled.  Now it is easily deduced,
that the transformation $\rm I\!R^8(\vec{u})\to\rm I\!R^5(\vec{x})$
must take the form [6]
%====================================================================
\begin{eqnarray*}
\left(
\begin{array}{c}
x_0\\x_1\\x_2\\x_3\\x_4\\0\\0\\0
\end{array}
\right)=\left(
\begin{array}{cccccccc}
u_0&u_1&u_2&u_3&-u_4&-u_5&-u_6&-u_7\\
u_4&u_5&-u_6&-u_7&u_0&u_1&-u_2&-u_3\\
u_5&-u_4&u_7&-u_6&-u_1&u_0&-u_3&u_2\\
u_6&u_7&u_4&u_5&u_2&u_3&u_0&u_1\\
u_7&-u_6&-u_5&u_4&u_3&-u_2&-u_1&u_0\\
u_1&-u_0&u_3&-u_2&u_5&-u_4&u_7&-u_6\\
u_2&-u_3&-u_0&u_1&-u_6&u_7&u_4&-u_5\\
u_3&u_2&-u_1&-u_0&-u_7&-u_6&u_5&u_4\\
\end{array}
\right)
\left(
\begin{array}{c}
u_0\\u_1\\u_2\\u_3\\u_4\\u_5\\u_6\\u_7
\end{array}
\right).
\end{eqnarray*}
%====================================================================

Hence it follows that
%==================================(5)==============================
\begin{eqnarray}
x_0&=&u_0^2+u_1^2+u_2^2+u_3^2-u_4^2-u_5^2-u_6^2-u_7^2,
\nonumber\\[2mm]
x_1&=&2 \,(u_0 u_4-u_1 u_5-u_2 u_6-u_3 u_7),
\nonumber\\[2mm]
x_2&=&2 \,(u_0 u_5+u_1 u_4-u_2 u_7+u_3 u_6),
\\[2mm]
x_3&=&2 \,(u_0 u_6+u_1 u_7+u_2 u_4-u_3 u_5),
\nonumber\\[2mm]
x_4&=&2 \,(u_0 u_7-u_1 u_6+u_2 u_5+u_3 u_4).
\nonumber
\end{eqnarray}
%=====================================================================
It is easy to prove that for the matrix $H(u;8)$ there is a condition
%=====================================================================
\begin{eqnarray*}
H(u; 8)H^T(u;8)=u^2 E(8)
\end{eqnarray*}
%=====================================================================
that guarantee the validity of Euler identity.

Adding to (5) the transformations [9]
%======================================(6)==============================
\begin{eqnarray}
\alpha_T&=&\frac{i}{2}\, \ln \,\frac{(u_0+i u_1)(u_2-i u_3)}
{(u_0-i u_1)(u_2+i u_3)} \in [0, 2\pi),
\nonumber\\[2mm]
\beta_T&=&2 \arctan \left(\frac{u_0^2+u_1^2}{u_2^2+u_3^2}\right)^{1/2}
\in [0, \pi],
\\[2mm]
\gamma_T&=&\frac{i}{2}\, \ln \,\frac{(u_0-i u_1)(u_2-i u_3)}
{(u_0+i u_1)(u_2+i u_3)} \in [0, 4\pi),
\nonumber
\end{eqnarray}
%===========================================================================
we obtain a transformation converting $\rm I\!R^8$ to the direct product
$\rm I\!R^5 \otimes{\bf S^3}$ of the space  $\rm I\!R^5(\vec{x})$
and a three-dimensional sphere ${\bf S^3}(\alpha_T,\beta_T,\gamma_T)$.

\section{Dyon-Oscillator Duality}

In the coordinates (5), (6) we can transform the Sch\"odinger equation for
the eight-dimensional isotropic oscillator
%=============================================================
\begin{eqnarray*}
\frac{{\partial}^2\psi}{\partial u^2_\mu}
+ \frac{2m}{\hbar^2}\left(E-\frac{m\omega^2u^2}{2}\right)\psi=0,
\qquad u_{\mu}\in R^8
\end{eqnarray*}
%===============================================================
into the equation [10]
%==============================(7)======================================
\begin{eqnarray}
\frac{1}{2m}\left(-i\hbar \frac{\partial}{\partial x_j}-
\hbar A^a_j \hat T_a\right)^2\psi + \frac{\hbar^2}{2mr^2}
{\hat T}^2\psi - \frac{e^2}{r}\psi = \epsilon \psi
\end{eqnarray}
%====================================================================
where $\epsilon$ and $e^2$ are defined by the relation (3).
The operators ${\hat T}_a$ are the generators of the $SU(2)$ group. In
coordinates $(\alpha_T,\beta_T,\gamma_T)$ they are parametrized as follows
%=======================================================================
\begin{eqnarray*}
{\hat T}_1&=&
i\left(\cos\alpha_T\, \cos\beta_T\frac{\partial}{\partial \alpha_T}
+\sin\alpha_T\frac{\partial}{\partial \beta_T}
-\frac{\cos\alpha_T}{\sin\beta_T}\, \frac{\partial}{\partial \gamma_T}
\right), \\[2mm]
{\hat T}_2&=&
i\left(\sin\alpha_T\, \cot\beta_T\frac{\partial}{\partial \alpha_T}
-\cos\alpha_T\frac{\partial}{\partial \beta_T}
-\frac{\sin\alpha_T}{\sin\beta_T}\, \frac{\partial}{\partial \gamma_T}
\right), \\[2mm]
{\hat T}_3&=& -i\frac{\partial}{\partial \alpha_T}.
\end{eqnarray*}
%=========================================================================
Recall that the operators ${\hat T}_a$ satisfy the following
commutation relations
%=======================================================================
\begin{eqnarray*}
\left[{\hat T}_a, {\hat T}_b\right] = i \varepsilon_{abc}{\hat T}_c.
\end{eqnarray*}
%=========================================================================

The five-dimensional vectors $\vec{A}^a$ are given by the expressions
%=======================================================================
\begin{eqnarray*}
{\vec A}^1 &=& \frac{1}{r(r+x_0)}\,(0, x_4,  x_3, -x_2, -x_1), \\[2mm]
{\vec A}^2 &=& \frac{1}{r(r+x_0)}\,(0, -x_3, x_4, x_1, -x_2),  \\[2mm]
{\vec A}^3 &=& \frac{1}{r(r+x_0)}\,(0,x_2, -x_1, x_4,-x_3),
\end{eqnarray*}
%==========================================================================
The vectors ${\vec A}^a$ are orthogonal to each other,
%=======================================================================
\begin{eqnarray*}
A_j^a A_j^b=\frac{1}{r^2}\, \frac{r-x_0}{r+x_0}\,\delta_{ab}
\end{eqnarray*}
%==========================================================================
and to the vector $\vec{x}=(x_0, x_1, x_2, x_3, x_4)$ as well.

The equation (7) is identical to the Pauli equation and therefore
we can give to the triplet of five-dimensional vectors ${\vec A}^a$
the meaning of the vector potentials with the line of singularity along
the nonpositive $x_0$ semiaxis.

The five-dimensional vector potentials ${\vec B}^a$
%===================================================================
\begin{eqnarray*}
\vec B^1 &=& \frac{1}{r(r - x_0)}(0,- x_4, x_3, -x_2, x_1), \\ [2mm]
\vec B^2 &=& \frac{1}{r(r - x_0)}(0, -x_3, -x_4, x_1, x_2),  \\ [2mm]
\vec B^3 &=& \frac{1}{r(r - x_0)}(0, x_2, -x_1, -x_4, x_3)
\end{eqnarray*}
%===================================================================
with the singularity axis, directed along the nonnegative $x_0$ semiaxis,
are obtained from the vectors ${\vec A}^a$ by the following gauge
transformation
%===============================================================
\begin{eqnarray*}
B_j = {\hat S}A_j{\hat S}^{-1} +
i{\hat S}\frac{\partial}{\partial x_j}{\hat S}^{-1}.
\end{eqnarray*}
%===================================================================
Here $A_j=A^a_j{\hat T}_a$, $B_j=B_j^a{\hat T}_a$, and
%===============================================================
\begin{eqnarray*}
{\hat S} = e^{-i\gamma {\hat T}_3}e^{-i\beta {\hat T}_2}
e^{-i\alpha {\hat T}_3}.
\end{eqnarray*}
%===================================================================
The hyperspherical angles $\alpha$, $\beta$ and $\gamma$ are
defined as
%==================================================================
\begin{eqnarray*}
\alpha &=& \frac{i}{2}\ln{\frac{(x_2 - ix_1)(x_4 - ix_3)}
{(x_2 + ix_1)(x_4 + ix_3)}} \in [0, 2\pi)  \\ [2mm]
\beta  &=& 2\arctan{\left(\frac{x_1^2 + x_2^2}
{x_3^2 + x_4^2}\right)^{1/2}} \in [0, \pi] \\ [2mm]
\gamma &=& \frac{i}{2}\ln{\frac{(x_2 + ix_1)(x_4 - ix_3)}
{(x_2 - ix_1)(x_4 + ix_3)}} \in [0, 4\pi).
\end{eqnarray*}
%============================================================

Now, it is necessery to explain what the physical system the
equation (7) describes.

\section{Field Tensor}

For the first step we rewrite the five-dimensional vector potentials
${\vec A}^a$ in the following form
%===========================(8)==================================
\begin{eqnarray}
A_i^{a} = \frac{2ig}{r(r+x_0)}{\tau}_{ij}^a x_j.
\end{eqnarray}
%=============================================================
Here ${\tau}_{ij}^a$ are the  $5\times5$ matrices having the
following explicit form
%=============================================================
\begin{eqnarray*}
{\tau}^1 = \frac{1}{2}\left(\begin{array}{ccc}
0&0&0\\
0&0&-i{\sigma}^1 \\
0&i{\sigma}^1&0
\end{array} \right),\,\,\,\,\,\,
{\tau}^2 = \frac{1}{2}\left(\begin{array}{ccc}
0&0&0\\
0&0&i{\sigma}^3 \\
0&-i{\sigma}^3&0
\end{array} \right),\,\,\,\,\,\,
{\tau}^3 = \frac{1}{2}\left(\begin{array}{ccc}
0&0&0\\
0&{\sigma}^2&0 \\
0&0&{\sigma}^2
\end{array} \right),
\end{eqnarray*}
%=============================================================
where ${\sigma}^a$ are the Pauli matrices and the ${\tau}^a$
matrices satisfying $[{\tau}^a,{\tau}^b] = i{\epsilon}_{abc}{\tau}^c$.
It is obvious that ${\tau}^a_{ij} = -{\tau}^a_{ji}$ and, therefore,
the vectors ${\vec A}^a$ are orthogonal to ${\vec x}$. Moreover, for
the ${\tau}^a$ matrices the following relations occur
%==============================(9)===============================
\begin{eqnarray}
4{\tau}_{ij}^a{\tau}_{jk}^b =
{\delta}_{ab}\left({\delta}_{ik} - {\delta}_{i0}{\delta}_{k0}\right)+
2i{\epsilon}_{abc}{\tau}_{ik}^c,
\end{eqnarray}
%=============================================================
%============================(10)===============================
\begin{eqnarray}
{\epsilon}_{abc}{\tau}_{ij}^b{\tau}_{km}^c = \frac{i}{2}\biggl[
\left({\delta}_{i0}{\delta}_{k0} - {\delta}_{ik}\right){\tau}_{jm}^a-
\left({\delta}_{i0}{\delta}_{m0} - {\delta}_{im}\right){\tau}_{jk}^a+
\nonumber \\ [2mm]
+\left({\delta}_{j0}{\delta}_{m0} - {\delta}_{jm}\right){\tau}_{ik}^a-
\left({\delta}_{j0}{\delta}_{k0} - {\delta}_{jk}\right){\tau}_{im}^a\biggr].
\end{eqnarray}
%=============================================================

Now, using the definition of the Yang-Mills field tensor
%=============================================================
\begin{eqnarray*}
F_{ij}^a =\frac{\partial A_j^a}{\partial x_i} -
\frac{\partial A_i^a}{\partial x_j} +
{\epsilon}_{abc}A_i^{b(+)}A_k^{c(+)}
\end{eqnarray*}
%=============================================================
and the expressions (8) and (10) we can write the field tensor
$F_{ij}^a$ in a more explicit form
%=========================================================
\begin{eqnarray*}
F_{ij}^a = \frac{1}{r^2}\left[
\left(x_j+r{\delta}_{j0}\right)A_i^a -
\left(x_i+r{\delta}_{i0}\right)A_j^a -
2i{\tau}_{ij}^a\right].
\end{eqnarray*}
%=============================================================
The straightforward computation gives
%=========================================================
\begin{eqnarray*}
F_{ij}^aF_{jk}^b = \frac{1}{r^6}
\left(x_ix_k-r^2{\delta}_{ik}\right){\delta}_{ab}
+ \frac{1}{r^2}{\epsilon}_{abc}F_{ik}^c.
\end{eqnarray*}
%=============================================================
or
%============================(11)=============================
\begin{eqnarray}
F_{ij}^aF_{ij}^b = \frac{4}{r^4}{\delta}_{ab}.
\end{eqnarray}
%=============================================================

\section{Topological Charge}

It is convenient to perform the computation of the topological charge
in the five-di\-men\-sio\-nal hyperspherical coordinates, which we define
as
%==============================================================
\begin{eqnarray*}
x_0 &=& r\cos\theta,   \\ [2mm]
x_2 + ix_1 &=& r\sin \theta \sin \frac{\beta}{2}e^{i\frac{\alpha
-\gamma}{2}},            \\ [2mm]
x_4 + ix_3 &=& r\sin \theta \cos \frac{\beta}{2}e^{i\frac{\alpha +
\gamma}{2}},
\end{eqnarray*}
%================================================================
where $r \in [0, \infty)$, $\theta \in [0, \pi]$, $\alpha \in [0, 2\pi)$,
$\beta \in [0, \pi]$, $\gamma \in [0, 4\pi)$.

It is known that the components of the second rank tensor in the
different coordinate systems are interrelated by the formula
%==============================================================
\begin{eqnarray*}
{\bar f}_{ik} = \frac{\partial x_m}{\partial {\bar x}_i}
\frac{\partial x_n}{\partial {\bar x}_k}f_{mn}
\end{eqnarray*}
%================================================================
in our case ${\bar x}_0=r$, ${\bar x}_1=\theta$,
${\bar x}_2=\beta$, ${\bar x}_3=\alpha$, ${\bar x}_1=\gamma$.
The direct computation gives that $F_{rk}^a \equiv 0$,
where $k=r, \theta, \beta, \alpha, \gamma$ (the other components
see in the Appendix).

Using the explicit forms of the hyperspherical components $F_{ij}^a$
it is possible to verify that the field tensor $F_{ij}^a$ is a self-duality
%============================(12)=============================
\begin{eqnarray}
{^*F}^{a(+)\mu \nu} = F^{a(+)\mu \nu}, \qquad \mu,\nu=1,2,3,4.
\end{eqnarray}
%=============================================================
Further, using the definition of the topological charge
%=========================================================
\begin{eqnarray*}
q = \frac{1}{32\pi^2}\sum_{a=1}^3
\oint {^*F}^{a(+)\mu \nu}F_{\mu \nu}^{a(+)}dS
\end{eqnarray*}
%=============================================================
where $dS = \frac{r^4}{8}{\sin}^3\theta \sin \beta
d\theta d\beta d\alpha d\gamma$, and taking into account equations
of the self-duality (12) and orthogonality conditions (11)
we obtain that in our case $q= + 1$.

So, we see that the equation (7) describe the charge-dyon system
with $SU(2)$ Yang monopole [11], and Schr\"odinger equation for the
eight-dimensional isotropic oscillator and equation (7) are dual
to each other.

It is important to note the following fact
%=========================================================
\begin{eqnarray*}
q^a = \frac{1}{32\pi^2}\oint {^*F}^{a(+)\mu \nu}F_{\mu \nu}^{a(+)}dS =
\frac{1}{3},
\end{eqnarray*}
%=============================================================
i.e. $1/3$ topololigical charge correspond to each $a$-th component of
the gauge field $F_{ij}^a$.

\section{Hidden Symmetry}

Since, our obtained system is a non-Abelian extension of the
five-dimensional Coulomb problem, it is natural to try to
construct an analog of the Runge-Lenz vector for Yang-Coulomb
monopole by passing from $\rm I\!R^3(\vec{x})$ to $\rm I\!R^5(\vec{x})$
and taking account of the gauge field [12]. The first step was
made many years ago [13]
%=============================================================
\begin{eqnarray*}
{\hat M}_k = \frac{1}{2\sqrt \mu_0}
\left({\hat p}_i{\hat l}_{ik}+
{\hat l}_{ik}{\hat p}_i +
\frac{2\mu_0e^2}{\hbar}\frac{x_k}{r}\right),
\end{eqnarray*}
%=============================================================
where ${\hat p}_i=-i\hbar\partial /partial x_i$,
${\hat l}_{ij} = \frac{1}{\hbar}\left(x_i{\hat p}_j - x_j{\hat p}_i\right)$.
The second step can be realized by the substitution [11]:
%=============================================================
\begin{eqnarray*}
{\hat p}_i &\to& {\hat \pi}_i = {\hat p}_i -\hbar A_i^a{\hat T}_a, \\ [2mm]
{\hat l}_{ij} &\to& {\hat L}_{ij} =
\frac{1}{\hbar}\left(x_i{\hat \pi}_j - x_j{\hat \pi}_i\right) -
r^2F_{ij}^a{\hat T}_a.
\end{eqnarray*}
%=============================================================
The following fundamental commutation relations are valid
%=============================================================
\begin{eqnarray*}
[{\hat \pi}_i,x_j] = -i \hbar {\delta}_{ij}, \qquad
[{\hat \pi}_i,{\hat \pi}_j] = i {\hbar}^2F_{ij}^a{\hat T}_a.
\end{eqnarray*}
%================================================================
It is possible to verify that
%=============================================================
\begin{eqnarray*}
[{\hat L}_{ik}, x_j] = i {\delta}_{ij}x_k - i {\delta}_{kj}x_i, \qquad
[{\hat L}_{ik}, {\hat \pi}_j] = i {\delta}_{ij}{\hat \pi}_k -
i {\delta}_{kj}{\hat \pi}_i, \\ [2mm]
[{\hat L}_{ij},{\hat L}_{mn}] = i {\delta}_{im}{\hat L}_{jn} -
i{\delta}_{jm}{\hat L}_{in} - i {\delta}_{in}{\hat L}_{jm} +
i{\delta}_{jn}{\hat L}_{im},
\end{eqnarray*}
%=============================================================
i.e. ${\hat L}_{ij}$ are indeed the generators of the group
$SO(5)$ and $[{\hat H}, {\hat L}_{ij}] = 0$.

After some calculations, we have $[{\hat H}, {\hat M}_i] = 0$
which means that ${\hat M}_i$ is the fact analog of the
Runge-Lenz vector for Yang-Coulomb monopole. It can also
be shown that
%=========================================================
\begin{eqnarray*}
[{\hat L}_{ij},{\hat M}_k] = i{\delta}_{ik}{\hat M}_j -
i{\delta}_{jk}{\hat M}_i, \qquad
[{\hat M}_i,{\hat M}_k] = -2i\hat H{\hat L}_{ik}.
\end{eqnarray*}
%=============================================================
These commutation rules generalize relations known from the theory
of the Coulomb problem [14].

Finally, let us introduce the $6\times 6$ matrix
%=============================================================
\begin{eqnarray*}
\hat D =
\left(\begin{array}{cc}
{\hat L}_{ij}&-{\hat M'}_i \\
{\hat M'}_j&0
\end{array} \right),
\end{eqnarray*}
%=============================================================
where ${\hat M'}_i = \left(-2\hat H\right)^{-1/2}{\hat M}_i$.
The components ${\hat D}_{\mu \nu}$ (where $\mu,\nu=0,1,2,3,4,5$)
satisfy the commutation relations
%=============================================================
\begin{eqnarray*}
[{\hat D}_{\mu \nu},{\hat D}_{\lambda \rho}] = i {\delta}_{\mu \lambda}
{\hat D}_{\nu \rho} - i {\delta}_{\nu \lambda}{\hat D}_{\mu \rho} -
i {\delta}_{\mu \rho}{\hat D}_{\nu \lambda} +
i {\delta}_{\nu \rho}{\hat D}_{\mu \lambda},
\end{eqnarray*}
%=============================================================
i.e. ${\hat D}_{\mu \nu}$ are the generators of the group
$SO(6)$. Since $[{\hat H}, {\hat D}_{\mu \nu}]=0$, one concludes
that Yang-Coulomb monopole is provided by the $SO(6)$ group of
hidden symmetry.

For the continuous spectrum ($\epsilon > 0$) we have
%=========================================================
\begin{eqnarray*}
[{\hat {\widetilde M}}_i,{\hat {\widetilde M}}_k] = -i{\hat L}_{ik},
\end{eqnarray*}
%=============================================================
where ${\hat {\widetilde M}}_i = \left(2\hat H\right)^{-1/2}{\hat M}_i$,
and the group of the hidden symmetry is $SO(5,1)$.

\section {Energy Levels}

The Casimir operators for $SO(6)$ are [15]
%=============================================================
\begin{eqnarray*}
{\hat C}_2 &=& \frac{1}{2}{\hat D}_{\mu \nu}
{\hat D}_{\mu \nu}, \\ [2mm]
{\hat C}_3 &=& {\epsilon}_{\mu \nu \rho \sigma \tau \lambda}
{\hat D}_{\mu \nu}{\hat D}_{\rho \sigma}
{\hat D}_{\tau \lambda},  \\ [2mm]
{\hat C}_4 &=& \frac{1}{2}{\hat D}_{\mu \nu}{\hat D}_{\nu \rho}
{\hat D}_{\rho \tau}{\hat D}_{\tau \mu}.
\end{eqnarray*}
%=============================================================
According [16], the eigenvalues of these operators can be
taken as
%=============================================================
\begin{eqnarray*}
C_2 &=& {\mu}_1({\mu}_1+4) + {\mu}_2({\mu}_2+2) + {\mu}_3^2, \\ [2mm]
C_3 &=& 48({\mu}_1+2)({\mu}_2+1){\mu}_3, \\ [2mm]
C_4 &=& {\mu}_1^2({\mu}_1+4)^2 + 6{\mu}_1({\mu}_1+4)
+ {\mu}_2^2({\mu}_2+2)^2 + {\mu}_3^4 - 2{\mu}_3^2,
\end{eqnarray*}
%=============================================================
where ${\mu}_1$, ${\mu}_2$ and ${\mu}_3$ are positive integers or
half-integers and ${\mu}_1 \geq {\mu}_2 \geq {\mu}_3$.

Direct calculation lead to the representation
%===========================(13)==================================
\begin{eqnarray}
{\hat C}_2 &=& - \frac{e^4\mu_0}{2{\hbar}^2\hat H} +
2{\hat T}^2 - 4, \nonumber \\ [2mm]
{\hat C}_3 &=& 48\left(-\frac{\mu_0e^4}{2{\hbar}^2
{\hat H}}\right)^{1/2}{\hat T}^2, \\ [2mm]
{\hat C}_4 &=& {\hat C}_2^2 + 6{\hat C}_2 -
4{\hat C}_2{\hat T}^2 - 12{\hat T}^2 + 6{\hat T}^4. \nonumber
\end{eqnarray}
%=============================================================
From the equation we can obtain another expression for the
eigenvalue $C_4$
%=============================================================
\begin{eqnarray*}
C_4 = \left[C_2 - 2T(T+1)\right]^2 +
6\left[C_2 - 2T(T+1)\right] + 2T^2(T+1)^2
\end{eqnarray*}
%=============================================================
and calculate that
%===========================(14,15)==================================
\begin{eqnarray}
C_2 - 2T(T+1) = {\mu}_1({\mu}_1+4),     \\ [2mm]
{\mu}_2^2\left({\mu}_2 +2\right)^2 + {\mu}_3^4 - 2{\mu}_3^2 =
2T^2(T+1)^2.
\end{eqnarray}
%=============================================================
The energy levels of the Yang-Coulomb monopole can be derived
from (13) and (14)
%==============================(16)================================
\begin{eqnarray}
{\epsilon}_{N}^T=-\frac{\mu_0e^4}{2\hbar^2(\frac{N}{2}+2)^2},
\end{eqnarray}
%====================================================================
where ${\mu}_1 = N/2$ and $N$--nonnegative integer number.
The substitution of the eigenvalues of ${\hat H}$ and ${\hat T}^2$
in the equation for ${\hat C}_3$ gives one more formula for $C_3$
%==============================================================
\begin{eqnarray*}
{C}_3 = 48({\mu}_1+2)T(T+1).
\end{eqnarray*}
%====================================================================
Now, we have two expressions for $C_3$ and the comparison leads the
relation
%===========================(17)===================================
\begin{eqnarray}
T(T+1) = ({\mu}_2+1){\mu}_3.
\end{eqnarray}
%====================================================================
Comparing (17) with (14) we will get the equation
%==============================================================
\begin{eqnarray*}
\left({\mu}_2^2 - {\mu}_3^2\right)
\left[({\mu}_2 +2)^2 - {\mu}_3^2\right] = 0.
\end{eqnarray*}
%====================================================================
Since ${\mu}_3 \leq {\mu}_2$, one concludes that ${\mu}_3={\mu}_2$.
Then, from (17) it follows that ${\mu}_2=T$. Therefore, $N$
in the formula (16) takes only values
$\frac{N}{2}=T, T+1, T+2,...$, -- the result known from our paper [17].

\section{Conclusions}

Formulae (5) and (6) together with the ansatz (3) form the
duality transformation mapping of the eight-dimensional
quantum oscillator into the charge-dyon system with the
$SU(2)$ Yang monopole. This type of duality is valid not only
for the 8D, 4D and 2D oscillators, but also for the
oscillator-like systems with the potentials
%=============================================================
\begin{eqnarray*}
V\left(u^2\right) = c_0 + c_1u^2 + W\left(u^2\right),
\end{eqnarray*}
%=============================================================
where $W\left(u^2\right)$ has a polynomial form
%=============================================================
\begin{eqnarray*}
W\left(u^2\right) = \sum_{n=2}^{\infty} c_n u^{2n}.
\end{eqnarray*}
%=============================================================
For such modified potentials, the ansatz (3) can be rewritten as
%=============================================================
\begin{eqnarray*}
\epsilon = - \frac{c_1}{4}, \qquad e^2 = \frac{E-c_0}{4}.
\end{eqnarray*}
%=============================================================

The hidden symmetry of the $SU(2)$ Yang-Coulomb monopole makes possible
to solve the equation (7) in the five-dimensional hyperspherical,
parabolic and elliptic coordinates by the separation of variables
method [17, 18, 19].

\vspace{5mm}

{\bf Acknowledgments}

I am very grateful to organizers who have invited me to this very
interesting Workshop. This work was 
partially supported  by the NATO Collaborative Linkage Grant No. 978431 and
ANSEF Grant No. PS81.

\section{Appendix}

1. The Cartesian components of the gauge field tensor $F_{ij}^a$:
%==============================================================
\begin{eqnarray*}
&& F_{01}^1 = -\frac{x_4}{r^3}, \qquad
F_{02}^1 = -\frac{x_3}{r^3}, \qquad
F_{03}^1 = \frac{x_2}{r^3}, \qquad
F_{04}^1 = \frac{x_1}{r^3}, \\ [1mm]
&&F_{12}^1 = \frac{x_2x_4-x_1x_3}{r^3(r+x_0)}, \qquad
F_{13}^1 = \frac{x_1x_2+x_3x_4}{r^3(r+x_0)}, \qquad
F_{14}^1 = \frac{1}{r^2}\left[\frac{x_1^2+x_4^2}{r(r+x_0)}-1\right], \\ [1mm]
&&F_{23}^1 = \frac{1}{r^2}\left[\frac{x_2^2+x_3^2}{r(r+x_0)}-1\right], \qquad
F_{24}^1 = \frac{x_1x_2+x_3x_4}{r^3(r+x_0)}, \qquad
F_{34}^1 = -\frac{x_1x_2+x_3x_4}{r^3(r+x_0)}.
\end{eqnarray*}
%================================================================

%==============================================================
\begin{eqnarray*}
&& F_{01}^2 = -\frac{x_3}{r^3}, \qquad
F_{02}^2 = \frac{x_4}{r^3}, \qquad
F_{03}^2 =- \frac{x_1}{r^3}, \qquad
F_{04}^2 = \frac{x_2}{r^3}, \\ [1mm]
&&F_{12}^2 = -\frac{x_1x_4+x_2x_3}{r^3(r+x_0)}, \qquad
F_{13}^2 = -\frac{1}{r^2}\left[\frac{x_1^2+x_3^2}{r(r+x_0)}-1\right],
\qquad
F_{14}^2 = \frac{x_1x_2-x_3x_4}{r^3(r+x_0)}, \\ [1mm]
&&F_{23}^2 = \frac{x_3x_4-x_1x_2}{r^3(r+x_0)}, \qquad
F_{24}^2 = \frac{1}{r^2}\left[\frac{x_2^2+x_4^2}{r(r+x_0)}-1\right], \qquad
F_{34}^2 = \frac{x_1x_4+x_2x_3}{r^3(r+x_0)}.
\end{eqnarray*}
%================================================================

%==============================================================
\begin{eqnarray*}
&& F_{01}^3 = -\frac{x_2}{r^3}, \qquad
F_{02}^3 = \frac{x_1}{r^3}, \qquad
F_{03}^3 =- \frac{x_4}{r^3}, \qquad
F_{04}^3 = \frac{x_3}{r^3}, \\ [1mm]
&&F_{12}^3 = \frac{1}{r^2}\left[\frac{x_1^2+x_2^2}{r(r+x_0)}-1\right],
\qquad
F_{13}^3 = \frac{x_2x_3-x_1x_4}{r^3(r+x_0)}, \qquad
F_{14}^3 = \frac{x_1x_3+x_2x_4}{r^3(r+x_0)}, \\ [1mm]
&&F_{23}^3 =- \frac{x_1x_3+x_2x_4}{r^3(r+x_0)}, \qquad
F_{24}^3 = \frac{x_2x_3-x_1x_4}{r^3(r+x_0)} , \qquad
F_{34}^3 =\frac{1}{r^2}\left[\frac{x_3^2+x_4^2}{r(r+x_0)}-1\right].
\end{eqnarray*}
%================================================================

2. The hyperspherical components of the gauge field tensor $F_{ij}^a$:
%==============================================================
\begin{eqnarray*}
&&F_{\theta \beta}^1 = \frac{1}{2}\sin \theta \sin \alpha,
\qquad
F_{\theta \alpha}^1 = 0, \qquad
F_{\theta \gamma}^1 = - \frac{1}{2}\sin \theta
\sin \beta \cos \alpha, \\ [1mm]
&&F_{\beta \alpha}^1 = - \frac{1}{4}{\sin}^2 \theta \cos \alpha, \qquad
F_{\beta \gamma}^1 = - \frac{1}{4}{\sin}^2 \theta
\cos \beta \cos \alpha, \qquad
F_{\alpha \gamma}^1 = \frac{1}{4}{\sin}^2 \theta
\sin \beta \sin \alpha.
\end{eqnarray*}
%================================================================

%=============================================================
\begin{eqnarray*}
&&F_{\theta \beta}^2 = \frac{1}{2}\sin \theta \cos \alpha, \qquad
F_{\theta \alpha}^2 = 0, \qquad
F_{\theta \gamma}^2 = \frac{1}{2}\sin \theta
\sin \beta \sin \alpha, \\ [1mm]
&&F_{\beta \alpha}^2 = \frac{1}{4}{\sin}^2 \theta \sin \alpha, \qquad
F_{\beta \gamma}^2 = \frac{1}{4}{\sin}^2 \theta
\cos \beta \sin \alpha, \qquad
F_{\alpha \gamma}^2 = \frac{1}{4}{\sin}^2 \theta
\sin \beta \cos \alpha.
\end{eqnarray*}
%=============================================================
%=============================================================
\begin{eqnarray*}
&&F_{\theta \beta}^3 = 0, \qquad
F_{\theta \alpha}^3 = \frac{1}{2}\sin \theta, \qquad
F_{\theta \gamma}^3 = \frac{1}{2}\sin \theta
\cos \beta, \\ [1mm]
&&F_{\beta \alpha}^3 = 0, \qquad
F_{\beta \gamma}^3 = -\frac{1}{4}{\sin}^2 \theta
\sin \beta, \qquad
F_{\alpha \gamma}^3 = 0.
\end{eqnarray*}
%=============================================================

\end{document}